\documentclass[floatfix,reprint,amsmath,amssymb,aps,pra]{revtex4-2}
\usepackage{xcolor}
\usepackage{graphicx}
\usepackage{dcolumn}
\usepackage{bm}
\usepackage[hidelinks]{hyperref}
\hypersetup{
     colorlinks   = true,
     citecolor    = blue,
     linkcolor    = blue
}

\usepackage{bbold}
\usepackage{yhmath}
\usepackage[english]{babel}
\usepackage[utf8]{inputenc}
\usepackage{xcolor}

\begin{document}

\preprint{APS/123-QED}

\title{Simultaneous photon and phonon lasing in a two-tone driven optomechanical system}
\author{Vitalie Eremeev$^{1,2,*}$, Hugo Molinares$^3$, Luis A. Correa$^{4,5}$, Bing He$^{6 }$}  
\email{vitalie.eremeev@udp.cl}
\email{bing.he@umayor.cl}

\affiliation{$^{1}$Instituto de Ciencias B\'asicas, Facultad de Ingenier\'ia y Ciencias, Universidad Diego Portales, Av.
 Ejercito 441, Santiago, Chile}
\affiliation{$^{2}$Institute of Applied Physics, Moldova State University, Academiei 5, MD-2028, Chi\c{s}in\u{a}u, Moldova} 
\affiliation{$^3$Departamento de Ciencias F\'{\i}sicas, Universidad de La Frontera, Casilla 54-D, Temuco, Chile}
\affiliation{$^4$Instituto Universitario de Estudios Avanzados (IUdEA), Universidad de La Laguna, La Laguna 38203, Spain}
\affiliation{$^5$Departamento de F\'isica, Universidad de La Laguna, La Laguna 38203, Spain}
\affiliation{$^6$Multidisciplinary Center for Physics, Universidad Mayor, Camino la Piramide 5750, Huechuraba, Santiago, Chile}

\begin{abstract}
Achieving simultaneous lasing of photons and phonons in optomechanical setups has great potential for applications in quantum information processing, high precision sensing and the design of hybrid photonic--phononic devices. Here, we explore this possibility with an optomechanical system driven by a two-tone field. Whenever the difference between the driving frequencies matches the associated mechanical frequency, the photon and phonon 
populations are found to achieve steady-state coherent oscillations, demonstrating a dual lasing phenomenon. Such drive--tone resonance condition can synchronize the phases of the photon and phonon fields, which facilitates a robust simultaneous lasing. Here, we provide analytical insights into the joint amplification of the optical and mechanical modes, and further confirm the dual lasing phenomenon by numerically calculating the relevant correlation functions and the power spectrum. Our setup, consisting of a single optomechanical cavity, is simpler than previous realizations of dual lasing and provides a clean picture of the underlying mechanisms. Our work thus paves the way for the development of novel strategies for the optimisation of optomechanical interactions through tailored driving schemes.
\end{abstract}

\maketitle

\section{Introduction}
\label{introduction}

Cavity optomechanics, which investigates the interactions between light and mechanical vibration within optical cavities, has advanced substantially in recent years. This interdisciplinary field bridges quantum optics and nanomechanics and has found applications in high-precision sensing, quantum information processing, as well as in the exploration of fundamental quantum mechanics at macroscopic scales \cite{Aspelmeyer2014, Kippenberg2008, Teufel2011, javid2021cavity, Chan2011,Montenegro2017,flor2023parametrically}. One of the key elements in quantum optics is lasing. While traditional (optical) lasing is well understood \cite{Scully97, Haken1984}, under suitable conditions it can also be extended to phonons---the quantized vibrational modes of a mechanical resonator \cite{Wallentowitz1996, Vahala2009, Grudinin2010,asano2016observation,sheng2020self,pan2024ultra}. In recent years, one of the most fascinating phenomena explored in cavity optomechanics is the possibility of \textit{simultaneous} lasing of photons and phonons \cite{Xiong2023, Fu2023, Wang2023}.

In a typical optomechanical system (OMS), the interaction between optical and mechanical modes is mediated by radiation pressure, which can either excite or damp the mechanical vibrations. For example, an OMS operating in a blue-detuned regime ($\Delta=\omega_c-\omega_l \approx -\omega_m$) under the rotating wave approximation (RWA), is dominated by an effect of \textit{two-mode squeezing}, that amplifies both mechanical and optical modes \cite{Aspelmeyer2014,lin2020entangling,he2023dynamical}. Here, $\omega_l$ is the pump frequency, $\omega_c$, the resonant frequency of the optical cavity, and $\omega_m$, the mechanical frequency of the system. Under suitable conditions this can induce a coherent amplification, i.e., a lasing effect. On the other hand, in a red-detuned regime ($\Delta=\omega_c-\omega_l\approx \omega_m$) the effective interaction reduces to a \textit{beam-splitter} effect that can be used for optomechanical cooling \cite{Wilson-Rae2007,marquardt2007quantum, schliesser2008resolved,Teufel2011,Chan2011,peterson2016laser, he2017radiation,wang2019breaking,lai2020nonreciprocal,liu2022accelerated} and quantum state transfer \cite{de2016quantum,weaver2017coherent,navarathna2023continuous, molinares2022high,molinares2023transfer}.
In contrast to the single-mode setting, the different scenarios emerging in an OMS driven by a two-tone laser field are much less understood \cite{shomroni2019two,Xiong2023,Wang2023}. Recent work by Xiong \textit{et al.} \cite{Xiong2023} experimentally explored the potential of a silicon-based optomechanical crystal cavity to achieve combined phonon and photon lasing; photon lasing was attributed to coherent scattering processes, which can, in principle, compete with and complement the Brillouin laser-like systems \cite{Otterstrom2018}. Likewise, Wang \textit{et al.} \cite{Wang2023} reported a two-domain laser based on the stimulated Brillouin scattering effect, mediated by long-lived flexural acoustic waves in a two-mode silica fiber ring cavity.

Here we focus on the driven optomechanical systems, for which the frequency difference between two drive tones matches the mechanical resonance frequency. This condition can enhance the optomechanical coupling. Specifically, the effective interaction between optical and mechanical modes can be modulated to realize locked oscillations of both cavity field and mechanical oscillator \cite{he2020mechanical, wu2022amplitude}. We show, within a fully quantum approach, that the resonance condition facilitates an efficient energy exchange between the optical and mechanical modes, thus creating optimal conditions for simultaneous lasing of photons and phonons. In order to capture both coherent interactions and dissipative processes, we resort to a quantum master equation to model the open dynamics and steady state properties of our optomechanical system  \cite{Gardiner2004, Clerk2010}. We thus have access to  lasing thresholds, second-order coherence, photon and phonon statistics, and the linewidth of the resulting lasing modes.

We apply both analytical and numerical methods to study the phenomenon. The analytical solution provides insights into the parameter regimes that support dual lasing, while our numerics offer a detailed view of the system dynamics and confirms the existence of stable lasing solutions. By determining the lasing threshold, we are able to identify the critical driving strengths and detuning conditions necessary to achieve lasing. Furthermore, we evaluate the second-order correlation function \( g^{(2)}(0) \), characterizing the degree of coherence of the lasing modes \cite{Scully97,Orszag2024} and the linewidth of the lasing modes, a key indicator of the coherence of the emitted field. We find that the latter is influenced by both intrinsic system properties and external driving conditions. Narrow linewidth is essential for applications requiring high coherence, such as precision metrology and quantum information processing. Our results suggest that, under optimal conditions, the linewidth of both photon and phonon lasing modes can be minimized, thus enhancing performance.

Simultaneous lasing of photons and phonons opens up the new possibilities for the development of hybrid optomechanical devices. For instance, high-precision sensors could leverage both optical and mechanical resonances, achieving unprecedented sensitivity, and hybrid quantum information systems could exploit the coherent interactions between photons and phonons for advanced data processing and communication protocols. Importantly, our findings contribute to a broader understanding of nonlinear dynamics and coherence in complex quantum systems, thus providing a foundation for future experimental realizations \cite{Vahala2008}.

This paper is structured as follows. In Sec.~\ref{sec2}, we introduce the conceptual model of the hybrid optomechanical system. In Sec.~\ref{sec3}, we explore the key indicators of phonon and photon lasing, including the analytical derivation of the stationary photon and phonon average and the effects of optomechanical coupling on the lasing threshold. We further study the second- and third-order coherence functions and the photon/phonon statistics and the power emission spectrum in Sec.~\ref{sec4}. Finally, in Sec.~\ref{sec6} we summarize and conclude.

\begin{figure}[t]
\centering
\includegraphics[width=0.95\linewidth]{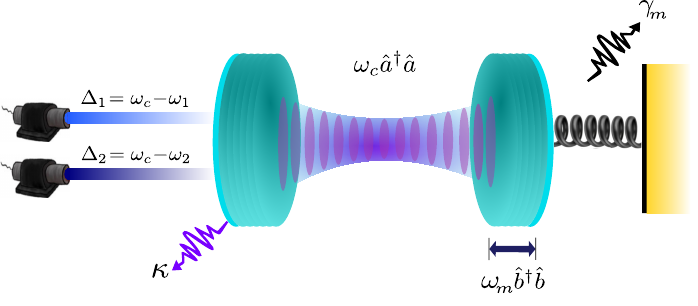}
\caption{An OMS driven by two pump fields, with their frequency difference being resonant with the mechanical frequency of the resonator.}
\label{fig1}
\end{figure}

\section{Two-tone driven--dissipative OMS}
\label{sec2}
Let us consider the OMS in Fig.~\ref{fig1}. The strength of the interaction between the cavity, of frequency $\omega_{c}$, and the mechanical oscillator (MO), of frequency $\omega_{m}$, is proportional to the coupling constant $g$. In the reference frame rotating at $\omega_{c}$ the Hamiltonian reads ($\hbar=1$)
\begin{multline} \label{eq:hamiltonian}
\mathcal{\hat{H}}=\omega_{m}\hat{b}^{\dagger}\hat{b}-i g\,\hat{a}^{\dagger}\hat{a}\,(\hat{b}^{\dagger}-\hat{b})\\
+ i\sum_{j=1}^2 E_{j}\left(\hat{a}^{\dagger}e^{i\Delta_{j}t} - \hat{a}\,e^{-i\Delta_{j}t} \right),
\end{multline}
where $\omega_{j}$ are the frequencies of the two-tone laser, $\Delta_{j}=\omega_{c}-\omega_{j}$ stands for the corresponding detunings, $E_{j}$ denotes the drive amplitudes, and $\hat{a}$ ($\hat{a}^{\dagger}$) and $\hat{b}$ ($\hat{b}^{\dagger}$) are the annihilation (creation) operators in the cavity and the MO, respectively. 

\begin{figure*}[t]
\centering
\includegraphics[width=0.26\linewidth]{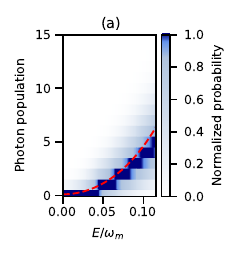}
\includegraphics[width=0.26\linewidth]{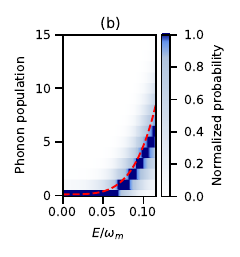}
\includegraphics[width=0.26\linewidth]{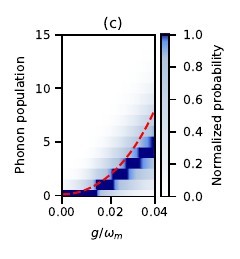}
\caption{Steady-state photon and phonon normalised occupation probabilities for $\Delta_{1}=\omega_{m}$ and $\Delta_{2}=0$. The probabilities obtained numerically from Eq.~\eqref{eq:master-equation} with the full (time-dependent) Hamiltonian in \eqref{eq:hamiltonian} are plotted as a color gradient. The dashed red curves correspond to the average steady-state occupations \eqref{eq:steady-state-occupations} obtained from the effective Hamiltonian $\hat{\mathcal{H}}_\text{eff}$ in Eq.~\eqref{eq:effective-H}. The photonic and phononic lasing thresholds can be clearly observed. Panels (a) and (b) illustrating the lasing threshold effect as a function of the driving field amplitude $E$ for a fixed optomechanical coupling of $g=0.03\,\omega_{m}$. Simultaneous photon-phonon lasing does occur above the threshold of $E_\text{cr}\sim 0.07\,\omega_{m}$ four our parameters. Conversely, panel (c) investigates the lasing threshold as a function of the optomechanical coupling $g$ for fixed $E=0.1\,\omega_{m}$. A phonon lasing threshold at $g\sim0.015\,\omega_{m}$ is observed, whereas there is no a photonic threshold for varying optomechanical coupling (not shown). The rest of parameters are $\kappa=0.1$, and $\gamma_{m}=6\times10^{-3}$, together with $\bar{n}_{c}=\bar{n}_{m}=0.1$ (all are expressed in units of $\omega_{m}$).}
\label{fig2}
\end{figure*}

In order to model the dissipation caused by the coupling to the surrounding environment we resort to a Markovian quantum master equation (ME) for the joint state $\hat{\rho}$ of cavity and mechanics in the rotating frame; namely,
\begin{multline}
     \frac{d\hat{\rho}}{dt} = -i[\mathcal{\hat{H}},\hat{\rho}]+\frac{\kappa}{2}\left(1+\bar{n}_{c}\right)\mathcal{L}_{\hat{a}}[\hat{\rho}]+\frac{\kappa}{2}\bar{n}_{c}\,\mathcal{L}_{\hat{a}}^{\dagger}[\hat{\rho}]\\
    +\frac{\gamma_m}{2}\left(1+\bar{n}_{m}\right)\mathcal{L}_{\hat{b}}[\hat{\rho}]+\frac{\gamma_m}{2}\bar{n}_{m}\,\mathcal{L}^\dagger_{\hat{b}}[\hat{\rho}].
    \label{eq:master-equation}
\end{multline}
Here the notation $\mathcal{L}_{\hat{\mathcal{O}}}[\hat{\rho}]$ stands for the dissipation superoperator of a decay channel with the associated jump operator $\hat{O}$ \cite{bp}; specifically
\begin{equation}\label{eq:dissipator}
    \mathcal{L}_{\hat{\mathcal{O}}}[\hat{\rho}] = 2\,\hat{\mathcal{O}}\hat{\rho} \,\hat{\mathcal{O}}^{\dagger}-\hat{\mathcal{O}}^{\dagger}\hat{\mathcal{O}}\hat{\rho}-\hat{\rho}\,\hat{\mathcal{O}}^{\dagger}\hat{\mathcal{O}},
\end{equation}
and $\kappa$ ($\gamma_{m}$) is the decay rate of the optical (mechanical) mode, with $\bar{n}_{c}$ ($\bar{n}_{m}$) denoting the average number of quanta in the photon (phonon) thermal bath. We note how the dissipation is modelled as if it acted \textit{separately} on the cavity and the mechanics, in spite of their mutual optomechanical coupling; i.e., we use a \textit{local} master equation. Such choice may be rigorously justified for a sufficiently weak $g$ \cite{trushechkin2016perturbative,gonzalez2017testing,hofer2017markovian}.

\section{Photon and phonon lasing}
\label{sec3}
In this section we discuss how simultaneous photon--phonon lasing can be realized by pumping an OMS with a two-tone drive, and study the corresponding lasing thresholds under various conditions. 

\subsection{Effective photon--phonon interaction} 
\label{sec31}

In order to qualitatively study the dynamics of photon--phonon lasing, it is convenient to move into a suitable interaction picture, in which the Hamiltonian \eqref{eq:hamiltonian} becomes
\begin{equation}\label{h1}
    \hat{\mathcal{H}}'=i\sum_{j=1}^2 E_{j}\left(\hat{a}^{\dagger}e^{i\Delta_{j}t}e^{-i\hat{F}(t)}-\hat{a}e^{-i\Delta_{j}t}e^{i\hat{F}(t)}\right).
\end{equation}
Full details of the transformation are deferred to Appendix~\ref{AppendA}. Here, we have defined 
\begin{equation*}
    \hat{F}\equiv\frac{g}{\omega_{m}}\left(\hat{b}^{\dagger}\eta+\hat{b}\,\eta^*\right),
\end{equation*}
and $\eta\equiv e^{i\omega_{m}t}-1$. We further assume that the optomechanical coupling $g$ is much smaller than the mechanical frequency $\omega_{m}$, which is consistent with typical experimental parameters as well as with our choice of master equation \eqref{eq:master-equation}. Hence,
$e^{-i \hat{F}}\simeq 1-i\,\hat{F}$ so that
\begin{equation*}
     \hat{\mathcal{H}}' \simeq i\sum_{j=1,2} E_{j}\left[\hat{a}^{\dagger}e^{i\Delta_{j}t}\left(1-i\,\hat{F}\right)\right. 
     \left.-\hat{a}\,e^{-i\Delta_{j}t}\left(1+i\,\hat{F}\right)\right],
\end{equation*}
In what follows, we consider the detuning combination $\Delta_{1} = \omega_{m}$ and $\Delta_{2} = 0$, which satisfies the resonance condition $\omega_1 - \omega_2 = \omega_{m}$ (swapping $\Delta_{1}$ for $\Delta_{2}$ would be equivalent). 
We take $\omega_j \gg \omega_m$ and $\omega_1 \simeq \omega_2$. Since the drive amplitudes are given by $E_j = \sqrt{2\kappa_{d} P/ \omega_j}$, with $\kappa_{d}$ measuring the coupling of the cavity to the driving field and $P$ denoting the pump power, we can further approximate $E_1 \simeq E_2 \equiv E$ (the case $ E_1 \neq E_2 $ is studied in Sec.~\ref{sec:amplitudes} below). Under these conditions, the resulting effective interaction leads to a two-mode squeezing effect for the photonic and phononic modes; namely, the transformed Hamiltonian $\hat{\mathcal{H}}'$ in the rotating-wave approximation becomes
\begin{equation}\label{eq:effective-H}
    \hat{\mathcal{H}}'\simeq\hat{\mathcal{H}}_\text{eff}\equiv i\,E\left(\hat{a}^{\dagger}-\hat{a}\right)-\frac{E\,g}{\omega_{m}}\left(\hat{a}^{\dagger}\hat{b}^{\dagger}+\hat{a}\,\hat{b}\right).
\end{equation}

The above effective Hamiltonian shows that the the resonance condition allows for the creation of photon-phonon pairs, closely analogous to the photon-photon pairs generated in parametric down-conversion \cite{Wu1986, Clerk2010}, which indicates a simultaneous photon and phonon amplification. A similar effective Hamiltonian can be found by means of a different linearization process \cite{lin2017mass, wang2019breaking, he2023dynamical}. In what follows, we  adopt $ \hat{\mathcal{H}}_\text{eff} $ merely for our \textit{qualitative} discussion, while the numerics supporting our main results are all based on the original nonlinear Hamiltonian in Eq.~\eqref{eq:hamiltonian}. 

\begin{figure*}[t]
\centering
\includegraphics[width=0.239\linewidth]{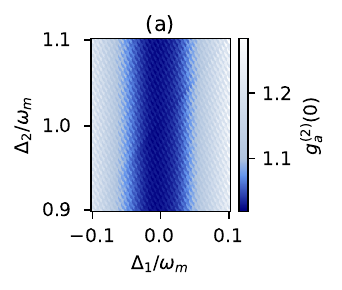}
\includegraphics[width=0.247\linewidth]{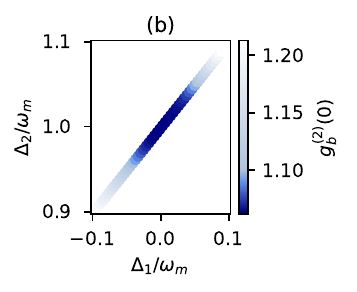}
\includegraphics[width=0.244\linewidth]{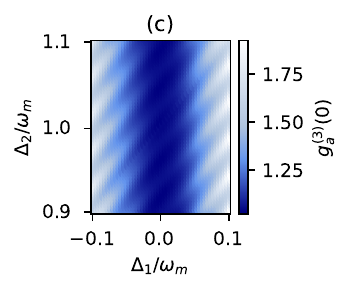}
\includegraphics[width=0.237\linewidth]{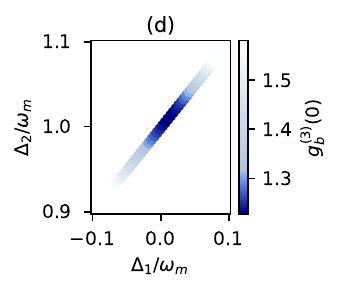}
\caption{Second-order and third-order correlation functions as a function of $\Delta_{1}$ and $\Delta_{2}$: cavity mode $(a, c)$, and mechanical mode $(b, d)$. All parameters are the same as in Figure ~\ref{fig2}, with $ g = 0.03\,\omega_m $ and $ E = 0.1\,\omega_m $. Simultaneous lasing only occurs close to the resonance condition $\Delta_1 = 0$ and $ \Delta_2 = \omega_m $. Just like in Figure ~\ref{fig2} we have used the master equation \eqref{eq:master-equation} with the full non-linear Hamiltonian $\hat{\mathcal{H}}$ in Equation ~\eqref{eq:hamiltonian}.}
\label{fig3}
\end{figure*}

\subsection{Steady-state output fields and lasing thresholds}
We can now evaluate the steady-state average photon and phonon numbers under the effective Hamiltonian in Eq.~\eqref{eq:effective-H}. To that end, we replace $\hat{\mathcal{H}}$ by $\hat{\mathcal{H}}_\text{eff}$ in Eq.~\eqref{eq:master-equation} and write the corresponding Heisenberg equations of motion for the closed set of dynamical variables $ \langle\hat{a}^\dagger\hat{a}\rangle $, $ \langle\hat{b}^\dagger\hat{b}\rangle $, $ \langle \hat{a}\hat{b} \rangle $, $\langle\hat{a}\rangle$ and $\langle \hat{b} \rangle$ (cf. Eqs.~\eqref{eq:heisenberg} in Appendix~\ref{AppendB}). These may then be solved for the steady-state occupations $\langle \hat{a}^\dagger\hat{a} \rangle_\infty$ and $\langle \hat{b}^\dagger\hat{b} \rangle_\infty$. The cumbersome resulting expressions \eqref{eq:steady-state-occupations} can be further simplified by neglecting the average thermal excitations $\bar{n}_c \simeq \bar{n}_m \simeq 0 $. This is realistic considering experimental conditions from, e.g., Refs.~\cite{Mir2020, Madiot2023}, where the mechanical mode had a frequency of $\omega_{m}/2\pi =6.8$~GHz. Thus, cooling the system to temperatures $T \sim 100$~mK would yield $\bar{n}_{m}\approx 10^{-2}$. With this in mind, the average steady-state occupations can be approximated as
\begin{subequations}\label{eq:ss_occupations}
\begin{align} \label{eq:ada_ss}
\langle\hat{a}^{\dagger}\hat{a}\rangle_\infty &=
\frac{{4 E^2 \gamma_m^2 \omega_{m}^2 \left[g^2\kappa + \omega_{m}^2 (\gamma_m + k)\right]}-16 E^4 g^4 \gamma_m }{{(\gamma_m + \kappa) ( \gamma_m\,\kappa\, \omega_{m}^2-4 E^2 g^2)^2}}  \\ 
\langle\hat{b}^{\dagger}\hat{b}\rangle_\infty &=
\frac{{4 E^2 g^2 \omega_{m}^2 \left[\gamma_m \kappa^2 + 4 E^2 (\gamma_m + \kappa)\right]}-16 E^4 g^4 \kappa }{{(\gamma_m + \kappa) ( \gamma_m\,\kappa\, \omega_{m}^2-4 E^2 g^2)^2}}  \label{eq:bdb_ss}
\end{align}
\end{subequations}
In Appendix~\ref{AppendC} we perform a stability analysis of the set of equations of motion \eqref{eq:heisenberg}, concluding that the above steady state can be attained under the condition 
\begin{equation}\label{eq:stability}
    E < \frac{\omega_m}{2g}\sqrt{\kappa\,\gamma_m}.
\end{equation}
Eq.~\eqref{eq:stability} is crucial from the experimental perspective, as it sets the limit of lasing stability.
 
From Eqs.~\eqref{eq:ss_occupations} one finds that in the absence of optomechanical coupling ($g = 0$), there is no steady-state phonon population and that the photon population scales as $\bar{n}_a \equiv \langle \hat{a}^\dagger\hat{a} \rangle_\infty \sim E^2$. In turn, for small optomechanical coupling, i.e., $g \ll \omega_{m}$, the average number of photons still scales as $\bar{n}_a \sim E^2$, while the average phonon occupation scales as $\bar{n}_b \equiv \langle \hat{b}^\dagger\hat{b} \rangle_\infty \sim E^4$. Therefore, given a weak pump field, the average number of phonons increases more slowly than the photon population does, leading to a higher threshold for phonon lasing; however, above threshold, the steady-state population of phonons can exceed the steady-state population of photons. 

\begin{figure*}[t]
\centering 
\includegraphics[width=0.39\linewidth]{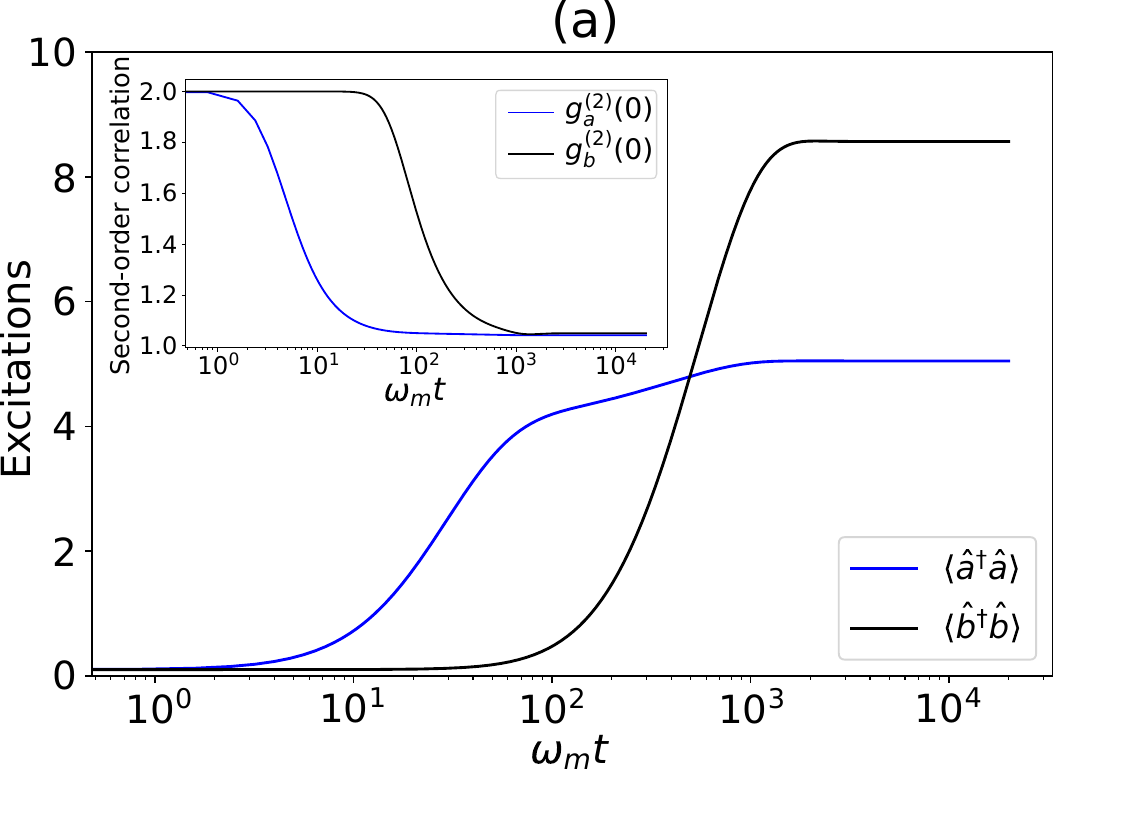}
\includegraphics[width=0.293\linewidth]{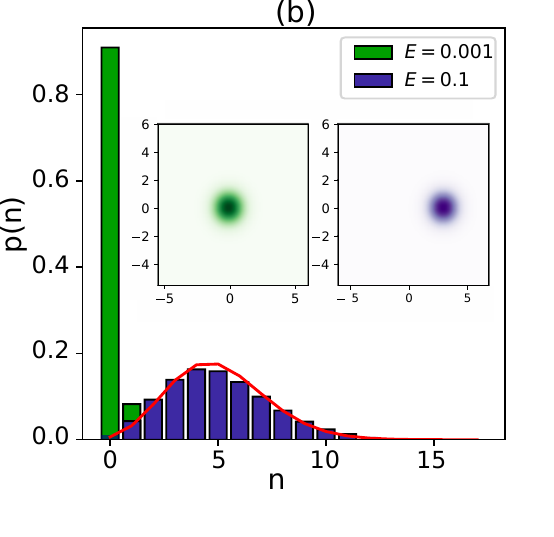}
\includegraphics[width=0.292\linewidth]{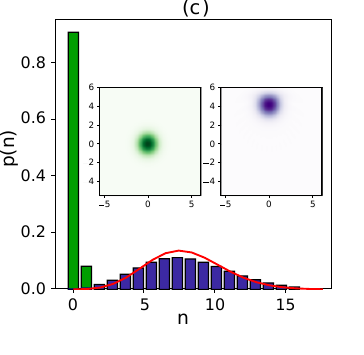}
\caption{(a) Time evolution of the average photon and phonon number above the lasing threshold. (inset) Dynamics of the second-order correlation function. The field strength is $E=0.1\,\omega_m$. Photon (b) and phonon (c) probability distributions below (green) and above threshold (purple). A Bose--Einstein distribution is observed below the lasing threshold ($E = 0.001\,\omega_{m}$), while a Poisson distribution is found above it ($E=0.1\,\omega_{m}$). This is compared with a standard Poisson distribution with the same mean (solid red line). (insets) Wigner functions of the resulting thermal (green) and coherent (purple) states. Other parameters (in units of $\omega_{m}$) are: $g=0.04$, $\kappa=0.1$, $\gamma_{m}=6\times10^{-3}$, $\bar{n}_{c}=\bar{n}_{m}=0.1$. All calculations use Eqs.~\eqref{eq:hamiltonian} and \eqref{eq:master-equation}.}
\label{fig4}
\end{figure*}

In Fig.~\ref{fig2}, we present the normalized population distribution for both photons and phonons drawn from the steady-state solutions of Eq.~\eqref{eq:master-equation}. To obtain them, we employed the numerical methods outlined in \cite{qutip}. For comparison, in Fig.~\ref{fig2} we also overlay the analytical expressions in Eqs.~\eqref{eq:steady-state-occupations} for the average steady-state occupations (red dashed lines) obtained from the effective Hamiltonian $\hat{\mathcal{
H}}_\text{eff} $ \eqref{eq:effective-H}; these are in very good agreement. Fixing the optomechanical coupling $g$, we observe a distinct lasing threshold for both photons and phonons at specific values $E_\text{cr}$ of the pump intensity (cf. Figs.~\hyperref[fig2]{2(a)} and \hyperref[fig2]{2(b)}). Beyond these, the system starts exhibiting Poissonian statistics, signalling the change from a thermal to a coherent state. 
Conversely, fixing the pump intensity and varying the coupling $ g $ does leaves photon lasing almost unaffected. This suggests that the phenomenon of photon lasing is predominantly influenced by the pump intensity and the precise tuning of the laser frequencies in relation to the optical cavity. This is further corroborated in Figs.~\ref{fig3}--\ref{fig5}. In contrast, as shown in Fig.~\hyperref[fig2]{2(c)} phonons do undergo stimulated amplification beyond a certain coupling threshold. This provides a control mechanism for phonon lasing. Indeed, comparing Figs.~\hyperref[fig2]{2(b)} and Fig.~\hyperref[fig2]{2(c)} we see that, as the optomechanical coupling increases, phonon lasing can be achieved at lower pump intensity.

\section{Photon--phonon statistics and power spectrum}
\label{sec4}

\subsection{Coherence}

The existence of lasing must be confirmed by the field autocorrelation functions \cite{Glauber1963} 
\begin{align*}
    g_{\mathcal{O}}^{(k)}(\tau)&=\frac{\langle\hat{\mathcal{O}}^{\dagger k}(t+\tau)\,\hat{\mathcal{O}}^{k}(t)\rangle}{\langle\hat{\mathcal{O}}^{\dagger}(t)\,\hat{\mathcal{O}}(t)\rangle^{k}}
\end{align*}
at zero time delay $\tau = 0$, with $\hat{\mathcal{O}} = \{\hat{a},\hat{b}\}$. Specifically, here we look into second- and third-order correlations ($k=2,3$), which quantify the degree of coherence in the emitted photons (and phonons), thus helping us to distinguish between thermal and coherent states. 

In Fig.~\ref{fig3} we plot the autocorrelation functions calculated from the steady-state solutions of Eqs.~\eqref{eq:hamiltonian} and \eqref{eq:master-equation} at various detunings. We have chosen a drive amplitude above the lasing threshold. Note that, away from the resonance $\Delta_{1} = 0$ and  $\Delta_{2}= \omega_{m}$, both photons and phonons exhibit autocorrelations consistent with thermal statistics (i.e., $g^{(k)}(0)\sim k!$). In contrast, around the resonance point $g^{(k)}(0)\to 1$, which indicates a coherent output and thus, genuine lasing \footnote{The same is true around $\Delta_1 = \omega_{m}$ and $\Delta_2 = 0$.}. Importantly, evidence of dual lasing may be observed over a \textit{range} $\Delta_1 = \pm \delta$ and $ \Delta_2 = \omega_m \pm \delta $. For instance, with our parameters $\delta\simeq 0.05\,\omega_m$, which provides sufficient measurement flexibility for experimental verification. 

In order to further illustrate the changes in the field states of the OMS, we plot the steady-state photon and phonon number distributions, both below and above threshold, as well as the corresponding Wigner quasi-probability distributions (cf. Figs.~\hyperref[fig4]{4(b)} and \hyperref[fig4]{4(c)}). Finally, in Fig.~\hyperref[fig4]{4(a)} we illustrate the dynamics of this lasing transition; namely, we plot the time evolution of the average photon and phonon numbers alongside the second-order autocorrelation function.

\subsection{Power spectrum}
\label{sec5}

The existence of a sharp peak in the power emission spectrum, with a linewidth much narrower than the natural linewidth of the system, is the hallmark of lasing. We thus set out to calculate the photon and phonon power spectra (PS) in the frequency domain numerically. The PS is given by a Fourier transform of the optical and mechanical second-order correlation functions $g^{(2)}_\mathcal{O}(\tau)$ in the steady-state regime \cite{Clerk2010}; namely, 
\begin{equation*}
    S_{\mathcal{O}}(\omega)\equiv\int_{-\infty}^{+\infty}d\tau\,e^{-i\omega\tau}\,\langle\hat{\mathcal{O}}^{\dagger}(\tau)\hat{\mathcal{O}}(0)\rangle_\infty,
\end{equation*}
where $\hat{\mathcal{O}}=\{\hat{a},\hat{b}\}$. By analyzing the photon and phonon's PS, we can identify the characteristic frequencies and linewidths associated with the optical and mechanical modes of the dual laser. This information is crucial to understand the energy transfer mechanisms and the interplay between the photonic and phononic subsystems. 

\begin{figure*}[t]
\centering
\includegraphics[width=0.98\linewidth]{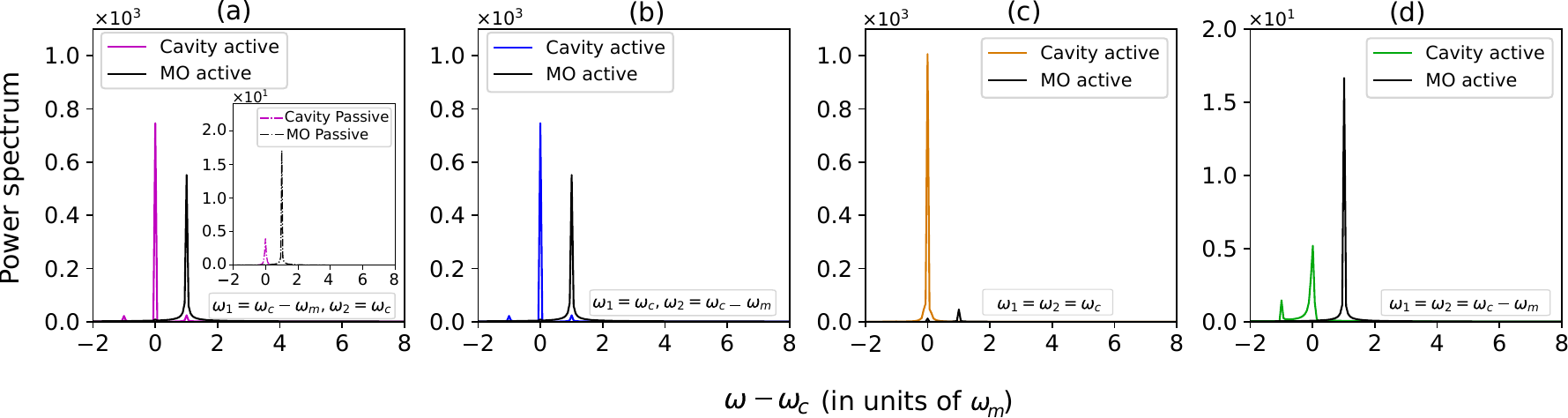}
\caption{Power spectrum of photons (colored) and phonons (black) under various resonance conditions for the drive frequencies (indicated on each panel) versus $ (\omega-\omega_c)/\omega_m $. Panel (a) depicts the active case ($E=0.1\,\omega_{m}$) alongside the passive case ($E=0.001\,\omega_m$), in the inset. Both (a) and (b) satisfy $|\Delta_{1}-\Delta_{2}|=\omega_{m}$ so that dual lasing is possible. Above threshold both cavity and mechanical modes exhibit a remarkable three-order-of-magnitude enhancement in their emission peaks. (c) Only photon lasing may be achieved for $\omega_1=\omega_2=\omega_c$, (d) no lasing is attained setting $\omega_1=\omega_2=\omega_c-\omega_m$. All other parameters are the same 
as in Fig.~\ref{fig3}.}
\label{fig5}
\end{figure*}

As illustrated in Fig.~\ref{fig5}, when the cavity frequency is chosen as $\omega_{c} \gg \omega_m$ the power spectra differ substantially depending on whether the system is below or above the lasing threshold (passive and active case, respectively). In the passive case (inset of Fig.~\hyperref[fig5]{5(a)}), the system is dominated by thermal photons and phonons, as evidenced by the low emission intensity. In contrast, in the active case a three-order-of-magnitude enhancement is observed at the corresponding emission peaks (see Figs.~\hyperref[fig5]{5(a)} and ~\hyperref[fig5]{5(b)}), thus demonstrating dual lasing whenever the resonance condition $|\Delta_1 - \Delta_2| = \omega_m$ is met. As shown in Figs.~\hyperref[fig5]{5(c)} and ~\hyperref[fig5]{5(d)}, operating out of resonance does not allow for dual lasing, even above threshold. 

Finally, note how the cavity mode exhibits multiple secondary peaks at integer multiples of $\omega_{m}$, albeit much smaller than the central one at $\omega=\omega_c$. This observation is consistent with the findings in Ref.~\cite{Nunnenkamp2014} for $\kappa \gg \gamma_{m}$.

\section{Discussion and conclusions}
\label{sec6}
In this paper we have demonstrated how a single optomechanical setup pumped with an appropriate two-tone driving field can attain dual photonic and phononic lasing. Crucially, the frequency-match condition $|\Delta_1-\Delta_2|=\omega_m$ is necessary for the joint amplification of phonon and photon fields. Under classical modelling, it is possible to show that these frequency combinations can indeed give rise to an amplitude- and phase-locking phenomenon for the mechanical mode \cite{he2020mechanical, wu2022amplitude}. Here, however, we have tackled the problem fully quantum-mecanically. In particular, we have qualitatively explained the dual lasing phenomenon, by deriving an effective Hamiltonian \eqref{eq:effective-H} for our system, which is dominated by a two-mode squeezing term, analogous to that of a linearized OMS under a single blue-detuned drive. Interestingly, such simple effective Hamiltonian can be used to obtain experimentally valuable expressions for the photonic/phononic lasing thresholds in terms of the driving amplitude and the optomechanical coupling. 

We have further confirmed that we deal with a genuine lasing scenario by numerically 
computing the higher-order autocorrelation functions and the power spectrum of photons and phonons around the resonance, which are completely based on the nonlinear dynamics. To that end, we have worked with the full system Hamiltonian and accounted for noise and dissipation through a Markovian quantum master equation. Although our main focus is on the driving tones with the same amplitude $E_1 = E_2 = E$, a more general case can also lead to dual lasing; see the discussion in Appendix~\ref{sec:amplitudes}.

Compared with the previous works of dual laser \cite{Xiong2023, Wang2023}, our main 
results are completely based on nonlinear dynamics (without any linearization except for an effective Hamiltonian used for the interpretation) and the setup is also much simpler. It is merely by driving a common optomechanical system with a two-tone field that is sufficiently strong. This is possible, for instance, with a system of suspended mechanical membranes \cite{sheng2020self}, which can usually withstand high pump powers. Interestingly, however, phononic lasing may be achieved not only by driving at high intensity, but also by increasing the optomechanical coupling $ g $ (see Fig.~\hyperref[fig2]{2(c)}). In fact, the relevant parameter to produce a coherent phononic output is the product $ g\,E$ (always ensuring that $ g\ll\omega_m $ holds) as illustrated in Fig.~\hyperref[fig4n]{6(c)}. A similar observation had been made in Ref.~\cite{lin2021catastrophic} in the context of phononic lasing.

The development of hybrid devices capable of dual lasing may pave the way towards a new generation of high-precision quantum sensors, such as optomechanical accelerometers or microresonators for bio-sensing \cite{javid2021cavity,flor2023parametrically}. Our results provide 
some practical insights for the implementation of such dual photon/phonon lasing on readily available optomechanical setups, as well as a neat physical picture of the underlying physical mechanisms, thus paving the way for future applications to quantum technologies.

\vspace{0cm}
\section*{Acknowledgements}
VE and BH acknowledge the financial support from Fondecyt Regular No. 1221250. HM acknowledge partial financial support from the project Code “FRO 21991”, from the Ministry of Education of Chile and Universidad de La Frontera. LAC acknowledges funding by the Spanish MICINN and the EU (FEDER) (PID2022-138269NB-I00*), and the Ram\'{o}n y Cajal Fellowship RYC2021325804-I funded by MCIN-AEI-10.13039-501100011033 and “NextGenerationEU”/PRTR. 

\bibliography{biblio.bib}

\appendix

\onecolumngrid

\section{\label{AppendA}Derivation of the effective Hamiltonian}

In this appendix, we present the derivation of the Hamiltonian in the interaction picture, which serves as the foundational model for this study. We begin by calculating the Hamiltonian in the first interaction picture
\begin{equation}\label{cuadro1}
    \mathcal{V}=e^{\imath \mathcal{H}_{0} t}\mathcal{H}_{1}e^{-\imath \mathcal{H}_{0} t},
\end{equation}
where $\mathcal{H}_{0}=\omega_{m}\hat{b}^{\dagger}\hat{b}$ and $\mathcal{H}_{1}=i g\hat{a}^{\dagger}\hat{a}\left(\hat{b}^{\dagger}-\hat{b}\right)+i\sum_{j=1,2} E_{j}\left(\hat{a}^{\dagger}e^{i\Delta_{j}t}-\hat{a}e^{-i\Delta_{j}t}\right)$.

From Eq. (\ref{cuadro1}), we readily get $\mathcal{V}=\mathcal{V}_{0}+\mathcal{V}_{1}$, where
\begin{align}
    \mathcal{V}_{0}&=-i g\hat{a}^{\dagger}\hat{a}\left(\hat{b}^{\dagger} e^{i\omega_{m} t}-\hat{b} e^{-i\omega_{m} t}\right)\equiv\hat{a}^{\dagger}\hat{a}\hat{f}(t),\\
    \mathcal{V}_{1}&=i\sum_{j=1,2} E_{j}\left(\hat{a}^{\dagger}e^{i\Delta_{j}t}-\hat{a}e^{-i\Delta_{j}t}\right),
\end{align}
with $\hat{f}(t)=-ig\left(\hat{b}^{\dagger}e^{i\omega_{m}t}-\hat{b}e^{-i\omega_{m}t}\right)$.
Next, we move to a second interaction picture, which is defined as follows
\begin{equation}
    \mathcal{V}'=\exp{\left\{i\int\mathcal{V}_{0}dt\right\}}\mathcal{V}_{1}\exp{\left\{-i\int\mathcal{V}_{0}dt\right\}}.
\end{equation}
This transformation allows us to readily obtain
\begin{equation}
    \mathcal{V}'=i\sum_{j=1,2} E_{j}\left(\hat{a}^{\dagger}e^{i\Delta_{j}t}e^{-i\hat{F}(t)}-\hat{a}e^{-i\Delta_{j}t}e^{i\hat{F}(t)}\right),
\end{equation}
with the Hermitian operator $\hat{F}(t)=\frac{g}{\omega_{m}}\left(\hat{b}^{\dagger}\eta+\hat{b}\eta^{*}\right)$ and $\eta=e^{i\omega_{m}t}-1$.  

In the following, we assume that the optomechanical coupling strength $g$ is significantly smaller than the mechanical frequency $\omega_{m}$, allowing for the following consideration: $e^{-i \hat{F}(t)}\approx1-i\frac{g}{\omega_{m}}\left(\hat{b}^{\dagger}\eta+\hat{b}\eta^{*}\right)$. Therefore
\begin{equation}\label{h1q}
    \mathcal{V}''=i\sum_{j=1,2} E_{j}\left[\hat{a}^{\dagger}e^{i\Delta_{j}t}\left(1-i\frac{g}{\omega_{m}}(\hat{b}^{\dagger}\eta+\hat{b}\eta^{*})\right)
     -\hat{a}e^{-i\Delta_{j}t}\bigg(1+i\frac{g}{\omega_{m}}(\hat{b}^{\dagger}\eta+\hat{b}\eta^{*})\bigg)\right].
\end{equation}
Now, by considering $E=E_{j}$ and the regime $\Delta_{1}=\omega_{m}$ and $\Delta_{2}=0$, the above equation leads to (keeping only the time independent terms):
\begin{equation}
    \mathcal{H}_\text{eff}=iE\left(\hat{a}^{\dagger}-\hat{a}\right)-\frac{Eg}{\omega_{m}}\left(\hat{a}^{\dagger}\hat{b}^{\dagger}+\hat{a}\hat{b}\right).
\end{equation}
This Hamiltonian reveals that 
OMS generates photon-phonon pairs that are similar to the photon-photon pairs produced in the parametric down-conversion. 

\section{\label{AppendB}Calculation of the average photon and phonon numbers}
To obtain the average number of photons and phonons, one can solve numerically the master equation (\ref{eq:master-equation}) with the Hamiltonian (\ref{eq:effective-H}). Alternatively, from the master equation \ref{eq:master-equation}, one can derive the Heisenberg equations for the average moments of the bosonic operators in order to close the set of equations and solve analytically for the average number of photons and phonons. Therefore, the set of equations for the bosonic moments reads:
\begin{subequations}\label{eq:heisenberg}
\begin{align}
      \frac{d\langle\hat{a}^{\dagger}\hat{a}\rangle}{dt}&= E\left(\langle\hat{a}^{\dagger}\rangle+\langle \hat{a}\rangle\right)+i\frac{Eg}{\omega_{m}}\left(\langle\hat{a}^{\dagger}\hat{b}^{\dagger}\rangle-\langle \hat{a}\hat{b}\rangle\right)-\kappa\langle\hat{a}^{\dagger}\hat{a}\rangle+\kappa\bar{n}_{a},\label{B1}\\
      \frac{d\langle \hat{b}^{\dagger}\hat{b}\rangle}{dt}&=
      i\frac{Eg}{\omega_{m}}\left(\langle\hat{a}^{\dagger}\hat{b}^{\dagger}\rangle-\langle \hat{a}\hat{b}\rangle\right)-\gamma_{m}\langle\hat{b}^{\dagger}\hat{b}\rangle+\gamma_{m}\bar{n}_{b},\\
      \frac{d\langle\hat{a}\hat{b}\rangle}{dt}&=
      E\langle\hat{b}\rangle+i\frac{Eg}{\omega_{m}}\left(\langle\hat{b}^{\dagger}\hat{b}\rangle+\langle\hat{a}^{\dagger}\hat{a}\rangle+1\right)-\left(\frac{\kappa}{2}+\frac{\gamma_{m}}{2}\right)\langle\hat{a}\hat{b}\rangle,\label{B3}\\
      \frac{d\langle\hat{b}\rangle}{dt}&=
      i\frac{Eg}{\omega_{m}}\langle\hat{a}^{\dagger}\rangle-\frac{\gamma_{m}}{2}\langle\hat{b}\rangle,\\
      \frac{d\langle\hat{a}\rangle}{dt}&=E+i\frac{Eg}{\omega_{m}}\langle\hat{b}^{\dagger}\rangle-\frac{\kappa}{2}\langle\hat{a}\rangle. \label{B5}
\end{align}
\end{subequations}
By setting the left-hand side of the equations \ref{B1}-\ref{B5} and the Hermitian conjugates of Eqs. \ref{B3}-\ref{B5} to zero and solving the resulting system, we can derive the average numbers of photons and phonons in the steady-state: 
\begin{subequations}\label{eq:steady-state-occupations}
\begin{align}
\langle\hat{a}^{\dagger}\hat{a}\rangle&=
\frac{{4 E^2 g^2 \gamma_m \kappa (\gamma_m (1 + \bar{n}_{a} - \bar{n}_{b}) - 2 \kappa \bar{n}_{b}) \omega_{m}^2 + \gamma_m^2 (\gamma_m + \kappa) (4 E^2 + \kappa^2 \bar{n}_{b}) \omega_{m}^4 -16 E^4 g^4 (\gamma_m + \gamma_m \bar{n}_{a} - \kappa \bar{n}_{b})}}{{(\gamma_m + \kappa) (\gamma_m \kappa \omega_{m}^2-4 E^2 g^2)^2}},  \\ 
\langle\hat{b}^{\dagger}\hat{b}\rangle&=
\frac{{4 E^2 g^2 (4 E^2 (\gamma_m + \kappa) + \gamma_m \kappa (\kappa (1 - \bar{n}_{a} + \bar{n}_{b})-2 \gamma_m \bar{n}_{a})) \omega_{m}^2 + \gamma_m^2 \kappa^2 (\gamma_m + \kappa) \bar{n}_{a} \omega_{m}^4 -16 E^4 g^4 (\kappa - \gamma_m \bar{n}_{a} + \kappa \bar{n}_{b})}}{{(\gamma_m + \kappa) (\gamma_m \kappa \omega_{m}^2-4 E^2 g^2)^2}}.  
\end{align} 
\end{subequations}

\section{\label{AppendC}Stability analysis of the steady-state solutions}
The stability of the above steady-state solutions, i.e. operative lasing condition, is determined by analyzing the system of equations of the form $\frac{d\vec{x}}{dt}=M\vec{x}+c$. Here, we have defined a vector  $\vec{x}=\left(\hat{a}^{\dagger}\hat{a},\hat{b}^{\dagger}b,\hat{a}\hat{b},\hat{a}^{\dagger}\hat{b}^{\dagger},\hat{a}, \hat{a}^{\dagger},\hat{b},\hat{b}^{\dagger}\right)^{T}$
where $c$ is a constant vector derived from the set of Eqs. \ref{B1}-\ref{B5} and $M$ is a matrix constructed from the elements on the right-hand side of Eqs. \ref{B1}-\ref{B5}
\begin{equation}
M = \begin{pmatrix}
   -\kappa & 0 & -iE g/\omega_m & iE g/\omega_m & E & E & 0 & 0 \\
   0 & -\gamma_m & -iE g/\omega_m & iE g/\omega_m & 0 & 0 & 0 & 0 \\
   iE g/\omega_m & iE g/\omega_m & -(\gamma_m+\kappa)/2 & 0 & 0 & 0 & E & 0 \\
   -iE g/\omega_m & -iE g/\omega_m & 0 & -(\gamma_m+\kappa)/2 & 0 & 0 & 0 & E \\
   0 & 0 & 0 & 0 & -\kappa/2 & 0 & 0 & iE g/\omega_m \\
   0 & 0 & 0 & 0 & 0 & -\kappa/2 & -iE g/\omega_m & 0 \\
   0 & 0 & 0 & 0 & 0 & iE g/\omega_m & -\gamma_m/2 & 0 \\
   0 & 0 & 0 & 0 & -iE g/\omega_m & 0 & 0 & -\gamma_m/2
\end{pmatrix}.
\end{equation}
The steady-state is achieved when the real parts of all eigenvalues of the matrix are negative. We calculate the eigenvalues for the matrix $M$ as follows:
\begin{align}
    \lambda_{1,2}&=-\frac{1}{2}\left(\kappa+\gamma_{m}\right),\\
    \lambda_{3,4}&=\frac{-\omega_m\left(\kappa+\gamma_{m}\right)-\sqrt{16E^2g^2+\omega_m^2\left(\gamma_m-\kappa\right)^2}}{4\omega_m},\\
    \lambda_{5}&=\frac{-\omega_m\left(\kappa+\gamma_{m}\right)-\sqrt{16E^2g^2+\omega_m^2\left(\gamma_m-\kappa\right)^2}}{2\omega_m},\\
    \lambda_{6,7}&=\frac{-\omega_m\left(\kappa+\gamma_{m}\right)+\sqrt{16E^2g^2+\omega_m^2\left(\gamma_m-\kappa\right)^2}}{4\omega_m},\\
    \lambda_{8}&=\frac{-\omega_m\left(\kappa+\gamma_{m}\right)+\sqrt{16E^2g^2+\omega_m^2\left(\gamma_m-\kappa\right)^2}}{2\omega_m}.
\end{align}
By considering a set of parameters $\{\omega_m,g,\kappa,\gamma_m,E\}>0$ we can conclude that the stability condition ($\mathcal{R}e[\lambda_{i}]<0$)  is satisfied only if
\begin{equation}
    E<\frac{\omega_m\sqrt{\kappa\gamma_m}}{2g}.
\end{equation}

\begin{figure*}[t]
\centering
\includegraphics[width=0.30\linewidth]{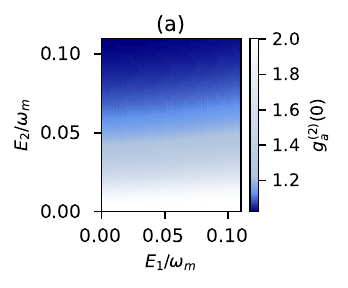}
\includegraphics[width=0.30\linewidth]{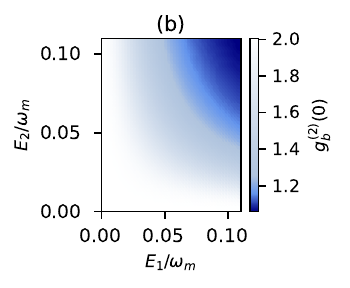}
\includegraphics[width=0.30\linewidth]{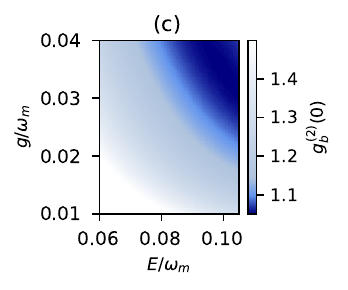}
\caption{Second-order correlation function as a function of $E_{1}$ and $E_{2}$ for the cavity (a) and the mechanical mode (b). All parameters are the same as in Figure ~\ref{fig3}. In panel (c) we show the onset of phononic lasing as a function of $E = E_1 = E_2$ and $ g $, illustrating that a sufficiently large \textit{product} $g\,E$ allows us to cross the lasing threshold.}
\label{fig4n}
\end{figure*}

\section{Coherence under a two-tone driving with $E_1 \neq E_2$}\label{sec:amplitudes}

For completeness, we now allow the amplitudes $E_1$ and $E_2$ to be different, while keeping $\Delta_1 = \omega_m$ and $\Delta_2 = 0$. In this case, the Hamiltonian \eqref{h1} under the rotating-wave approximation takes the form
\begin{equation}
     \mathcal{H}' = iE_2\left(\hat{a}^{\dagger}-\hat{a}\right) - \frac{g}{\omega_m}(E_1 - E_2)\left(\hat{a}\,\hat{b}^{\dagger} + \hat{a}^{\dagger}\,\hat{b}\right) - \frac{gE_2}{\omega_m}\left(\hat{a}^{\dagger}\hat{b}^{\dagger} + \hat{a}\,\hat{b}\right).
    \label{he2}
\end{equation}
Note that whenever $E_1 \neq 0$ and $E_2 = 0$, we are left with a beam-splitting term. Anti-Stokes scattering can then induce a mechanical cooling effect, or a state swap between the mechanical and the cavity mode. Lasing, however is not possible in this configuration, as shown in Figs.~\hyperref[fig4n]{6(a)} and ~\hyperref[fig4n]{6(b)}. 
Conversely, whenever $E_1=0$ and $E_2\neq0$, the Hamiltonian would describe a scenario where the Stokes and anti-Stokes scattering events induce amplification, while enabling state transfer between the mechanical and optical cavity modes. In this regime, the photon field can exhibit coherence beyond a threshold amplitude $E_2$ (cf. Fig.~\hyperref[fig4n]{6(a)}), but the phonon field is never coherent, as indicated by the autocorrelation function $g_b^{(2)}(0) > 1$ in Fig. \hyperref[fig4n]{6(b)}.

\end{document}